\documentstyle[11pt,newpasp,twoside,epsf]{article}

\begin{document}

 
\title{Nuclear Brightness Profiles of Merger Remnants:\\
Constraints on the Formation of Ellipticals by Mergers}

\author{Roeland P.~van der Marel and David Zurek}
\affil{Space Telescope Science Institute, Baltimore, USA}

	
\begin{abstract}
We present preliminary results of an HST/NICMOS program to image
merger remnants in the J, H and K bands. The nuclear brightness
profiles for most sample galaxies are similar to those typical for
elliptical galaxies, but some (including the well-studied NGC 3921 and
7252) have an unusually high luminosity density at small radii. This
is consistent with the prediction of N-body simulations that gas flows
to the center during a merger and forms new stars.
\end{abstract}

 
\keywords{galaxies: elliptical and lenticular, cD ---
          galaxies: individual (NGC~34, NGC~1316, NGC~3921, NGC~7252,
                                NGC~7727) ---
          galaxies: interactions ---
          galaxies: kinematics and dynamics ---
          galaxies: nuclei ---
          galaxies: structure.}


\section{Introduction}

The possibility that many elliptical galaxies formed from mergers of
disk galaxies is a topic of continuing interest. That mergers form
elliptical-like remnants has been demonstrated through numerical
simulations, and ground-based imaging has shown that many merger
remnants have r$^{1/4}$ luminosity profiles. These arguments, along
with the detection of shells, ripples and kinematically decoupled
cores in elliptical galaxies, support this `merger hypothesis' (e.g.,
Kennicutt, Schweizer, \& Barnes 1998; hereafter KSB98).

Theoretical arguments indicate that it is in the nuclei of remnants
where the merger hypothesis may face its most stringent test. If
dynamical relaxation is the dominant physical process in mergers, then
remnant nuclei will be very diffuse with large cores (Hernquist 1992),
unless the progenitor nuclei were dense to begin with. If both merging
galaxies contain a central black hole, then the stellar density of the
merger remnant will be lower than that of the progenitor galaxies
(Quinlan \& Hernquist 1997).  Alternatively, if mergers are
accompanied by strong gaseous dissipation and central starbursts, then
the remnant may have a high stellar density and steep luminosity
profile (Mihos \& Hernquist 1994).

A comparison between the observed nuclear properties of merger
remnants and elliptical galaxies can shed more light on the viability
of the merger hypothesis and on the physical processes that govern the
structure of merger remnants. The nuclear brightness profiles of
elliptical galaxies have been mapped in great detail with HST. Faber
et al.~(1997; hereafter F97) studied a large sample of normal
ellipticals. Carollo et al.~(1997; hereafter C97) studied a sample of
elliptical galaxies with kinematically decoupled cores (presumably old
merger remnants), and found few differences as compared to the sample
of F97. To complement this work, we initiated an HST study of a sample
of younger merger remnants (van der Marel, Zurek, Mihos, Heckman \&
Hernquist 2000, in preparation), and we present here some of the
preliminary results.

\section{Sample Selection}

A well-known compilation of nearby interacting galaxies and mergers is
Toomre's list of 11 galaxies selected from the NGC Catalog (see
KSB98). The two latest-stage mergers in the list are NGC 3921 and NGC
7252, which have tidal tails but show no remaining signs of two
galaxies with a separate identity. To create a sample for our study,
we sought galaxies with morphological properties similar to NGC 3921
and 7252 from the Catalogs of Arp (1966) and Vorontsov-Velyaminov
(1977), and from the imaging survey of (UV-bright) Markarian galaxies
by Mazzarella \& Boroson (1993). This yielded a sample of 19 galaxies
with $cz < 10000 \>{\rm km}\,{\rm s}^{-1}$, of which we imaged 14
galaxies (including NGC 3921 and 7252). The remaining five galaxies
are classified as ultra-luminous IR galaxies, and were imaged with HST
by other teams.


\begin{figure}
\epsfxsize=0.9\hsize
\centerline{\epsfbox{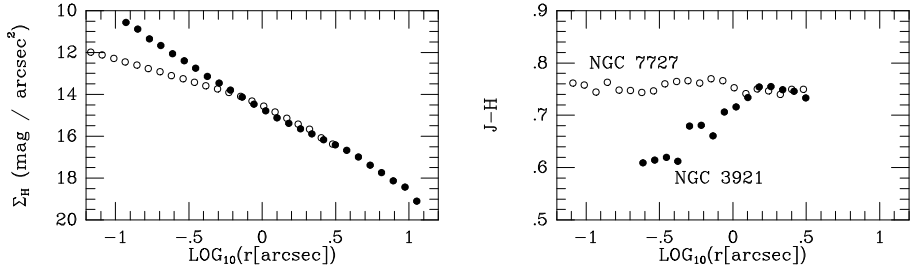}}
\caption{Solid dots show the observed H-band surface brightness
profile (left) and $J-H$ color (right) for NGC 3921. Open dots are for
NGC 7727, shifted horizontally to the distance of NGC 3921. NGC 7727
has a fairly typical `power-law' (F97) profile with $I(R) \propto
R^{-0.8}$ at the HST resolution limit. By contrast, NGC 3921 is
unusually blue and bright at small radii, with $I(R) \propto
R^{-1.9}$.}
\end{figure}

 
\section{Observations}

To minimize any influence of dust on the observed brightness profiles
we observed the galaxies in the near-IR with the HST/NICMOS instrument
(Cycle 7 project GO-7268). Images were obtained with the NIC2 camera
(pixel size $0.076''$ square) using the filters F110W, F160W and
F205W, corresponding roughly to J, H and K, and with the NIC1 camera
(pixel size $0.043''$ square) only in F110W.  Each image was subjected
to basic reduction steps followed by Lucy deconvolution with an
appropriate PSF. Azimuthally averaged brightness profiles were
extracted for all camera/filter combinations. Example results for two
galaxies are shown in Figure~1.

Each brightness profile was fit by a `nuker' law (Lauer et al.~1995),
which was deprojected to obtain the three-dimensional luminosity
density. Figure~2 shows this density in the $H$-band at a fiducial
radius $r=50 \>{\rm pc}$, as function of galaxy luminosity (assuming $H_0 =
80 \>{\rm km}\,{\rm s}^{-1} \> {\rm Mpc}^{-1}$), both for the galaxies
in our sample and for those in the samples of F97 and C97 (transformed
to the $H$-band under the assumption of a proto-typical $V-H = 3.0$
for elliptical galaxies; Peletier, Valentijn \& Jameson 1990; Silva \&
Bothun 1998). While most of the galaxies in our sample follow the same
approximate correlation as normal ellipticals, there are three
galaxies that strongly stand out because of their high luminosity
density: NGC~34, 3921 and 7252. One other galaxy that stands out for
the same reason is NGC 1316 from the F97 sample, which is also a
merger remnant (Schweizer 1980).

The $J-H$ and $H-K$ color profiles of NGC 3921 and 7252 show that they
become bluer towards the center, presumably due to recent star
formation. This is consistent with the detection of strong Balmer
absorption lines in ground-based spectra of these galaxies
(KSB98). NGC 34 becomes redder towards the center, probably as a
result of dust absorption. NGC 34 is the most IR-luminous galaxy of
those that we observed, suggesting the presence of ongoing or recent
star formation in this galaxy as well.


\begin{figure}
\epsfxsize=0.90\hsize
\centerline{\epsfbox{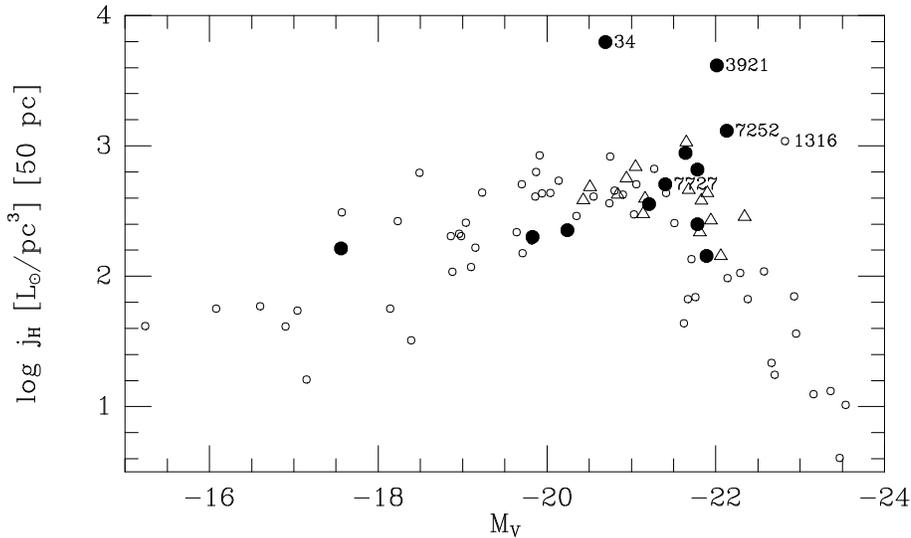}}
\caption{Three-dimensional $H$-band luminosity density at $r=50 \>{\rm
pc}$ as function of $M_V$. Solid dots are the galaxies in our
HST/NICMOS sample. Open dots and triangles are galaxies from F97 and
C97 respectively. NGC numbers are indicated for selected galaxies.}
\end{figure}

 
\section{Discussion and Conclusions}

The high luminosity densities observed in NGC~34, 3921 and 7252 are
probably a direct consequence of recent star formation triggered by a
merger. Stellar populations fade with time, and these galaxies will
therefore become more similar to normal ellipticals as time
passes. Dynamical and spectral evidence suggest that the mergers
happened $0.5$--$1.5$ Gyr ago (KSB98). The models of Bruzual \&
Charlot (e.g., 1993) indicate that a single-burst population fades by
a factor of $\sim 10$ between $0.5$ and 10 Gyr. Figure~2 therefore
suggests that these galaxies may become similar to normal ellipticals
within a Hubble time. Most galaxies in our sample are already now
similar to normal ellipticals in terms of their nuclear luminosity
density, although some fall on the high end of the range occupied by
normal ellipticals. If these galaxies are the remnants of disk-disk
mergers, then either the merger ages must be large so that the newly
formed stars have mostly faded, or they never formed many new stars,
e.g., because the progenitors were gas poor or the star formation
efficiency was low.

Results such as those for NGC 3921 in Figure~1 indicate that its star
formation was limited mostly to the central region, $r < 0.5'' \approx
200 \>{\rm pc}$. This is consistent with predictions of dissipative
N-body simulations of disk-disk mergers, in which the gas quickly
falls to the central few-hundred pc (Mihos \& Hernquist 1994). The
`excess' light in the central arcsec of NGC 3921 (as compared to NGC
7727; see Figure~1) represents $\sim 4$\% of the total galaxy
luminosity, and probably a smaller fraction in terms of mass. CO
complexes observed in the central kpc of merger remnants support the
view that gas flows to the center in galaxy interactions, but even if
all the CO observed in NGC 3921 and 7252 were soon turned into stars,
the light from recently formed stars would still provide only a small
fraction of the total galaxy luminosity (Hibbard \& Yun 1999).

To summarize, we have detected the luminosity spikes predicted by
dissipative simulations of disk-disk mergers, but only in some of our
galaxies. In general, it appears that the light from young stars does
not provide a major contribution to the total galaxy luminosity.  This
is consistent with work by Silva \& Bothun (1998), who found that the
near-IR colors of most morphologically disturbed ellipticals are
inconsistent with intermediate age (2--5 Gyr) stars providing much of
the luminosity. This raises the question whether these galaxies were
ever similar to ultra-luminous infrared galaxies, in which massive
starbursts are known to occur as a result of galaxy interactions.
 


\end{document}